\documentclass[twocolumn]{revtex4}
\usepackage{graphicx}

\makeatletter

\begin{document}

\title{Chaotic desynchronization of multi-strain diseases }

\author{Ira B.~Schwartz$^1$, Leah B.~Shaw$^1$, Derek
  A.~T.~Cummings$^3$,  Lora Billings$^2$, Marie
  McCrary$^2$, Donald S.~Burke$^3$}

\affiliation{$^1$US Naval Research Laboratory, Code 6792, Nonlinear Systems
  Dynamics Section, Plasma Physics Division, Washington, DC 20375}

\affiliation{$^2$Montclair State University, Department of Mathematical Sciences,
Upper Montclair, NJ 07043}

\affiliation{$^3$Johns Hopkins University, Department of International Health,
  Bloomberg School of Public Health, 615 N. Wolfe St., Baltimore, MD 21205}

\begin{abstract}
Multi-strain diseases are diseases that consist of several strains, or
serotypes. The serotypes may interact by antibody-dependent
enhancement (ADE), in which infection with a single serotype is
asymptomatic, but infection with a second serotype leads to serious
illness accompanied by greater infectivity. It has been observed from
serotype data of dengue hemorrhagic fever that outbreaks of the four
serotypes occur asynchronously. Both autonomous and seasonally driven
outbreaks were studied in a model containing ADE. For sufficiently
small ADE, the number of infectives of each serotype synchronizes,
with outbreaks occurring in phase. When the ADE increases
past a threshold, the system becomes chaotic, and infectives of each
serotype desynchronize. However, certain groupings  of the primary and secondary infectives remain synchronized even in the chaotic regime.
 
\end{abstract}

\maketitle

Recently there has been a wide range of work on chaotic synchronization in
dynamical systems \cite{manrubia,pikovsky}. When synchronizing chaotic
systems, almost all of the work deals with coupled or connected systems
\cite{BoccalettiKOVZ02} and analyzing their stability. 
In biological systems, such as population
models, synchronization may result from coupling strengths being enhanced
\cite{LloydJ04}, while desynchronization may take place as a result of
vaccine control as in measles \cite{RohaniEG99}. 
 { In this work, we consider the 
dynamics of a single population}
to shed light on the phase dynamics of multi-strain diseases 
\cite{BartleyDG02}.  { The dynamics observed exhibits phase-locked regular behavior, as well as chaotic phase desynchronization between strains.  Although we consider a single population model, we use the term ``synchronization'' to describe phase locking between variables \cite{footnote:cm}.}

 Many population models in the past have considered 
single strain diseases, as in childhood diseases. In
this case, the population may be grouped into the following
compartments: susceptibles, infectives, and recovered
\cite{Anderson91}. { With no seasonal forcing included in the model, the only endemic solution to the single
strain SIR models is an equilibrium point \cite{schwartz83}.}

 However, many diseases have co-circulating strains,
or serotypes, such as influenza \cite{AndreasenLL97}, malaria
\cite{GuptaTAD94}, and dengue virus \cite{FergusonDA99}. Such diseases
display anti-genic diversity, exhibiting distinct serotypes when
measured.  Recent efforts at modeling multi-strain diseases have
explored the oscillatory dynamics generated by multiple co-existing
serotypes with partial cross-immunity
\cite{Gog02,AndreasenLL97,EstevaV03}. However, current thinking
regarding the interacting serotypes of dengue virus is that
cross-reactive antibodies act to enhance the infectiousness of a
subsequent infection by another serotype \cite{Vaughn00}. This is
known as {\it antibody-dependent enhancement.}

It has been shown through recent serology measurements in Thailand
that dengue fever, which has four co-circulating serotypes, exhibits
asynchronous outbreaks. That is, each serotype has peaks that occur at
different times \cite{NisalakENKTSBHIV03}. { (See Fig.~\ref{fig:Nisalak data} in the Appendix.)} Note that most
observed infections are secondary \cite{NisalakENKTSBHIV03}, due to
increased symptom severity.

In this { paper}, we analyze how the antibody-dependent enhancement (ADE)
factor controls the onset of { oscillatory outbreaks}, as well as how asynchronous
secondary infections are controlled dynamically. 

\section{Description of model}

To model the spread of multi-strain diseases, we follow the approach
of Ferguson {\it et.~al} \cite{FergusonAG99}, where they restrict the
model to two serotypes. 
Our modeling approach differs in the general number of serotypes and in that all compartments are distinct from one another.
The full model for $n$ serotypes is described below. Our
simulations will be based on four serotypes, based on measured dengue data in
Thailand \cite{NisalakENKTSBHIV03}.  

The variable definitions are as follows: $s$, Susceptible to all
serotypes; $x_{i}$, Primary infectious with serotype $i$; $r_{i}$,
Primary recovered from serotype $i$; $x_{ij}$, Secondary infectious,
currently infected with serotype $j$, but previously had $i$ $(i\ne
j)$. The model is a system of ODEs describing the rates of change of
the population fractions within each compartment \cite{cummingsthesis}:
\begin{eqnarray}
\frac{ds}{dt} & = & \mu-\beta s\sum_{i=1}^{n}\left(x_{i}+\phi\sum_{j\ne i}x_{ji}\right)-\mu_{d}s\label{eq:model susceptibles}\\
\frac{dx_{i}}{dt} & = & \beta s\left(x_{i}+\phi\sum_{j\ne i}x_{ji}\right)-\sigma x_{i}-\mu_{d}x_{i}\label{eq: model primary}\\
\frac{dr_{i}}{dt} & = & \sigma x_{i}-\beta r_{i}\sum_{j\ne i}\left(x_{j}+\phi\sum_{k\ne j}x_{kj}\right)-\mu_{d}r_{i}\label{eq:model recovered}\\
\frac{dx_{ij}}{dt} & = & \beta r_{i}\left(x_{j}+\phi\sum_{k\ne j}x_{kj}\right)-\sigma x_{ij}-\mu_{d}x_{ij}.\label{eq:model secondary}\end{eqnarray}
The parameters $\mu$, $\mu_{d}$, $\beta$, and $\sigma$ denote birth, death, contact, and recovery rates, respectively.  { We assume that individuals who have recovered from two infections are immune to further infection since tertiary infections are reported very rarely \cite{NisalakENKTSBHIV03}.} The fixed parameters throughout the paper are given by: $\mu=\mu_{d}=0.02$, $\beta=200$, and $\sigma=100$, all with units of $years^{-1}$ \cite{cummingsthesis}. (Mortality rate, $\mu_{d}$, is set equal to the birth rate so that the population remains constant in time.) Antibody-dependent enhancement is governed by the parameter $\phi$, { which has not previously been measured for populations}. Notice that in Eqs.~\ref{eq:model susceptibles}-\ref{eq:model secondary}, ADE enters in a nonlinear enhancement factor when $\phi>1$. {We use a single $\phi$ for all strains for ease of analysis. Thus any loss of synchrony between the strains will result not from asymmetry but from the dynamics itself. Finally,} notice that since the value of $\mu$ is small compared to $\beta$ and $\sigma$, it can be considered as a small parameter.

\section{Results}

\subsection{Bifurcation structure}

\begin{figure}
\includegraphics[%
 width=2.75in,
 keepaspectratio]{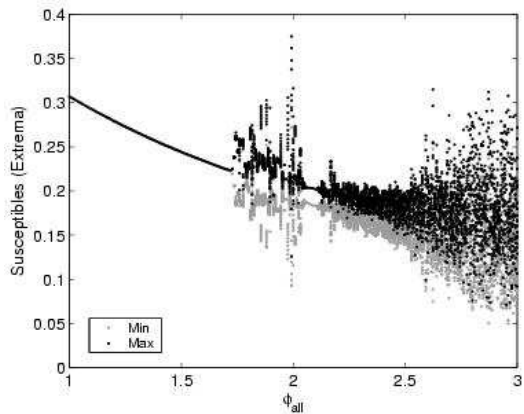} 
\caption{\label{fig:bifurcation diagram} {Bifurcation diagram for the autonomous model with  $\beta=200$, $\sigma=100$, $\mu=0.02$.  Shown for each $\phi$ are the maxima (black) and minima (gray) of the susceptibles during a 100 year run, after transients were removed.}}
\end{figure}

Unlike the usual SIR models for single strains, which in the absence of
forcing have only steady state behavior, the addition of multiple serotypes can
induce regular and chaotic outbreaks. In particular, for a critical value of
$\phi$, there exists a Hopf bifurcation to periodic oscillations. 
{ See the bifurcation diagram given in Fig.~\ref{fig:bifurcation diagram} for the
transition from steady state to oscillatory behavior as a function of $\phi$.} 
The usual trivial steady state, which has the population consisting of all susceptibles
$(s=1)$ and the rest of the components at zero, is unstable. 
(The trivial solution of Eqs.~\ref{eq:model susceptibles}-\ref{eq:model
  secondary} with $n=4$ serotypes has 4 unstable, 12 strongly attracting, and 5 weakly
attracting directions.) The non-trivial, or endemic,  steady state may be computed
numerically for arbitrary $\phi$. At steady state, we notice the following:
1. The primary infectives are equal. 2. The recovered variables are
equal. 3. All secondary infectives are equal. 

Compartmental equality at steady state holds before the Hopf bifurcation point as well as after the Hopf point (although past the Hopf bifurcation point the steady state solution is unstable). We make these assumptions about the model at equilibrium, and the resulting local dynamics can be reduced to a four-dimensional system: 
\begin{eqnarray}
\frac{dy_{1}}{dt} &=& \mu-n\,\beta\, y_{1} y_{2}-n\,(n-1)\,\beta\,\phi\, y_{1} y_{4}\label{eq:ADEreduced_sus} \\
\frac{dy_{2}}{dt} &=& \beta\,y_{1} y_{2}+(n-1)\,\beta\,\phi\,  y_{1}y_{4}-\sigma\, y_{2}\label{eq:ADEreduced_inf} \\
\frac{dy_{3}}{dt} &=& \sigma\, y_{2}-(n-1)\,\beta\, y_{3} y_{2}-(n-1)^2\,\beta\,\phi\, y_{3} y_{4}\label{eq:ADEreduced_rec} \\
\frac{dy_{4}}{dt} &=& \beta\,y_{3} y_{2}+(n-1)\, \beta\,\phi\,y_{3} y_{4}-\sigma\, y_{4}\label{eq:ADEreduced_sec}
\end{eqnarray} 
Notice that for simplicity we have removed the mortality terms in each of the variables, since they are of $\mathcal{O}(\mu)$ and have a negligible effect on the steady states. Moreover, removing the mortality terms allows an analytical estimate of the endemic steady states and stability. Mortality does need to be included in the long time asymptotic runs, which we do below. The reduced model has the following steady state solution: 
\begin{equation}
\left[ \frac{ \sigma}{ \beta \, \left(\phi+1\right)}, \frac{\mu}{n \, \sigma}, \frac{\sigma}{(n-1) \, \beta \, \left(\phi+1\right)}, \frac{\mu}{n\,(n-1) \, \sigma} \right]. \label{eq:ss_reducedmodel}
\end{equation}
 
Given the steady state solution as a function of $\phi$ in
Eq.~\ref{eq:ss_reducedmodel}, to compute the stability we need to evaluate the
linearization about the steady state. Therefore, we take the Jacobian of the
vector field of the reduced model about the steady state and examine the
characteristic polynomial for the eigenvalues. Recalling that $\mu$ is a small
parameter, we can expand the solutions to the characteristic equation in terms
of $\mu$. Since the data in
\cite{NisalakENKTSBHIV03} displays four serotypes, the number of serotypes is set to $n=4$.  We have a strongly attracting direction given by $z_{1}(\mu)=-\sigma+O(\mu)$, a weakly attracting direction given by $z_{2}(\mu)=-( 3 \,\beta \,\mu \,(\phi+1)^{2})/(4 \, \sigma)+O(\mu^{2})$, and a pair of complex eigenvalues: 
\begin{equation}
z_{\pm}(\mu)={\frac{ \beta \, D(\phi)}{8 \, \sigma}}\mu\,\pm \, i(\beta\mu)^{3/2}(1-f(\phi))+\mathcal{O}(\mu^{2}),
\label{eq:evpm}\end{equation}
where $f\prime(\phi)<0$. The sign of the expression $D(\phi) \equiv
3\,{\phi}^{2}-4\,\phi-4$ determines the stability of the complex pair. Since
$\phi$ is assumed to be greater than or equal to unity, $D(\phi)<0$ if $\phi
\in [1,2)$ and positive otherwise. Therefore, the steady state
  undergoes a Hopf bifurcation at $\phi = 2$. The results are close to
  numerical simulation, since mortality terms were dropped in the
  analysis but included in the simulations. 

\begin{figure}
 \hspace*{-3.2in} (a) \\ \vspace*{-0.1in}

  \includegraphics[%
 width=3in,
 keepaspectratio]{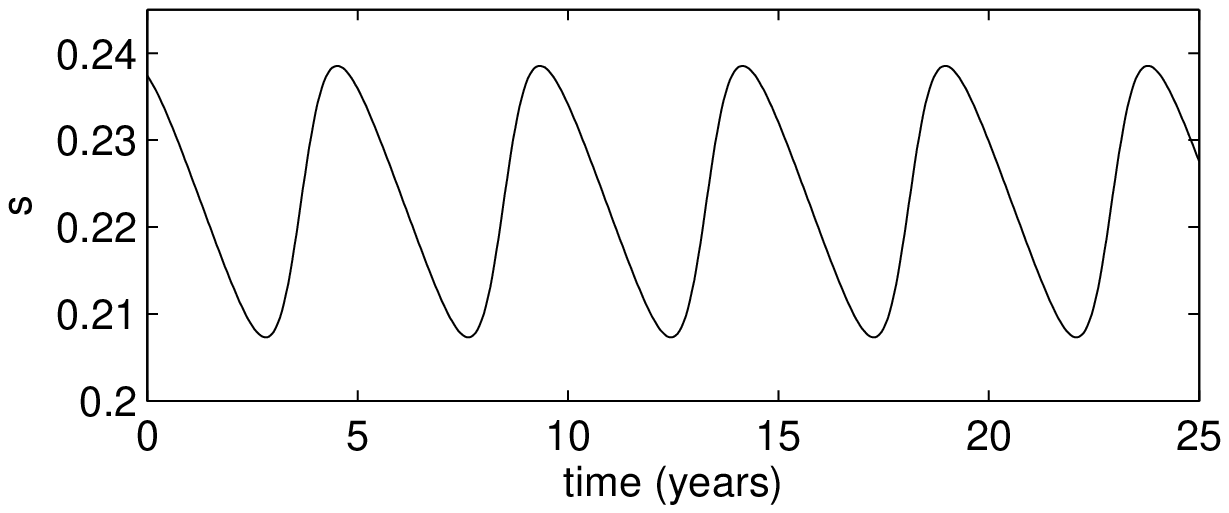} \\
 \hspace*{-3.2in} (b) \\ \vspace*{-0.1in}

 \includegraphics[%
 width=3in,
 keepaspectratio]{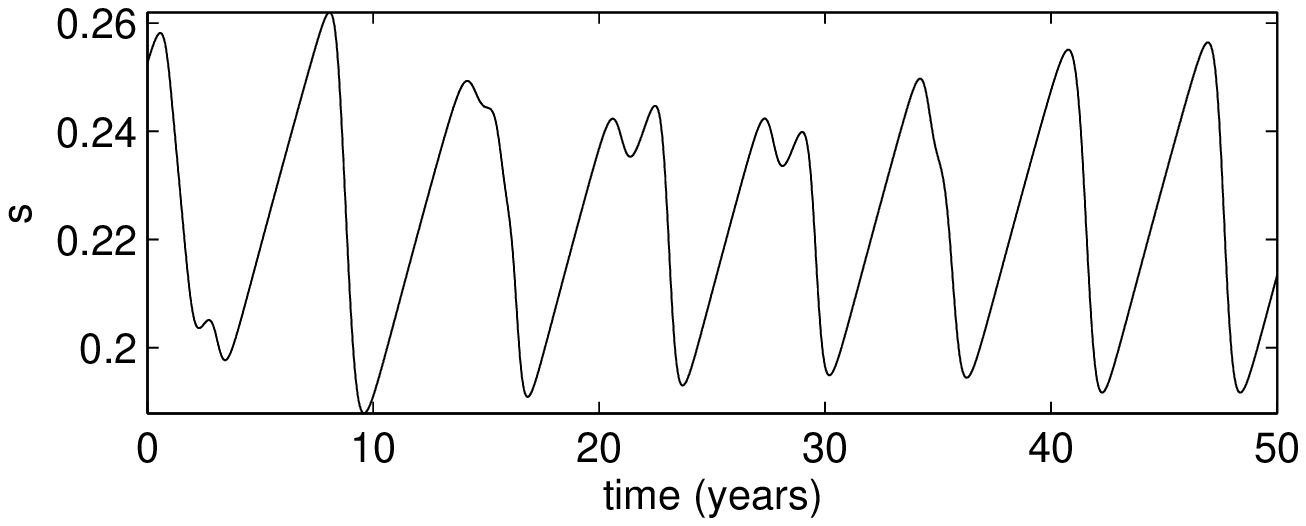} 

 \hspace*{-3.2in} (c) \\ \vspace*{-0.1in}

 \includegraphics[%
 width=3in,
 keepaspectratio]{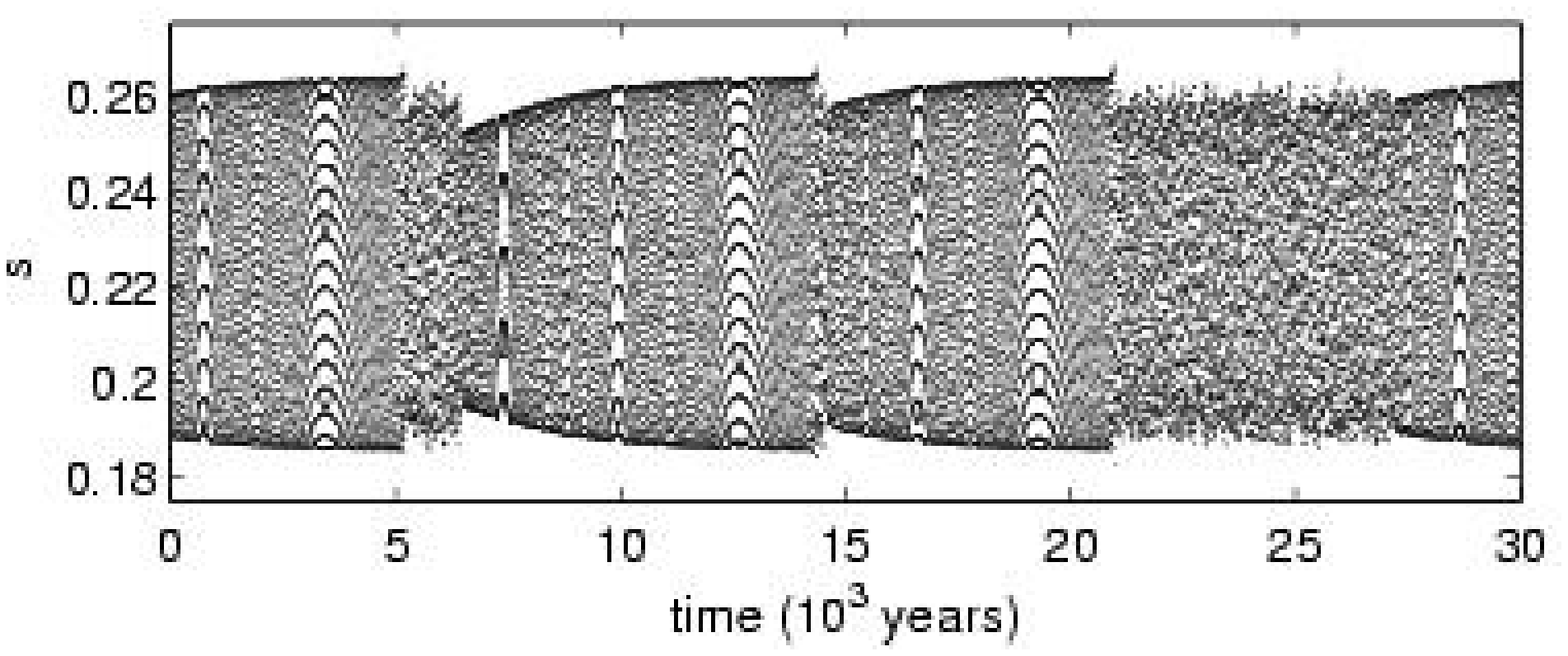} 
\caption{\label{fig:autonomous composite time series} Time series plots of the susceptibles for the autonomous case, where $\beta=200$, $\sigma=100$, $\mu=0.02$. (a) Periodic case for ADE factor $\phi=1.725$. { (b) Chaotic case for $\phi=1.73$. (c) Longer time series for chaotic case, $\phi=1.73$, with sampling once a year.} }
\end{figure}

{ Since the reduced model does not capture the asynchronous behavior past the Hopf point, we continue our analysis 
using the full model.} 
As we increase $\phi$ beyond the Hopf point, the dynamics exhibits periodic time series, as plotted in
Fig.~\ref{fig:autonomous composite time series}(a). The susceptibles exhibit a
period of approximately five years when $\phi=1.725$. 
{ The actual range of stable periodic solution is quite small, and occurs over a $\Delta \phi$ of 0.004838. 
(See the quick transition to irregular oscillations after the steady state in the bifurcation diagram given in 
Fig.~\ref{fig:bifurcation diagram}.) Past the $\phi$ value where
periodic solutions become unstable, we find chaotic behavior,
indicated by a positive maximum Lyapunov exponent for most $\phi$
values in that region. (The chaotic attractors persist over many
initial conditions chosen from a random distribution.) We note that in
this complicated region, there are small windows with attracting limit
cycles resulting in a zero maximum Lyapunov exponent. Lyapunov
exponents were computed by integrating linear variational equations
along solutions to  Eqs.~\ref{eq:model susceptibles}-\ref{eq:model
  secondary}. We show examples of chaotic oscillations in
Fig.~\ref{fig:autonomous composite time series} (b) and (c) for short
and long time series. Notice that in Fig.~\ref{fig:autonomous
  composite time series}(c)}, the time series exhibits oscillatory
regions which have a slowly growing envelope, interspersed with
chaotic intervals. We will exploit this structure to examine how each
serotype behaves dynamically.

\subsection{Phase analysis}

Since the measured data for dengue fever shows that the serotypes oscillate out
of phase, we investigate the phase of primary and secondary infectives with
respect to a particular secondary infective in the full model of Eqs.~\ref{eq:model
  susceptibles}-\ref{eq:model secondary}. To measure phase differences with
respect to a reference infective, let $Y(t)$ denote the reference infective, and
$Z(t)$ another infective. Let $\{t_{k}\}$ denote the sequence of times for  local
maxima of $Y(t)$, and $\{\tau_{k}$\} the times for local maxima of $Z(t)$. For $\tau_{m}
\in (t_{k},t_{k+1})$, define the phase of $Z$ relative to $Y$ in the interval
as $\Psi_{ZY}(\tau_{m})=2\pi\frac{\tau_{m}-t_{k}}{t_{k+1}-t_{k}}$.  

\begin{figure} [tp] \hspace*{-3in} (a) \\
\vspace*{0.9in} \hspace*{-3in} (b) \\
\vspace*{0.9in} \hspace*{-3in} (c) \\
\vspace*{-2.25in} 

\includegraphics[%
 width=2.75in,
 keepaspectratio]{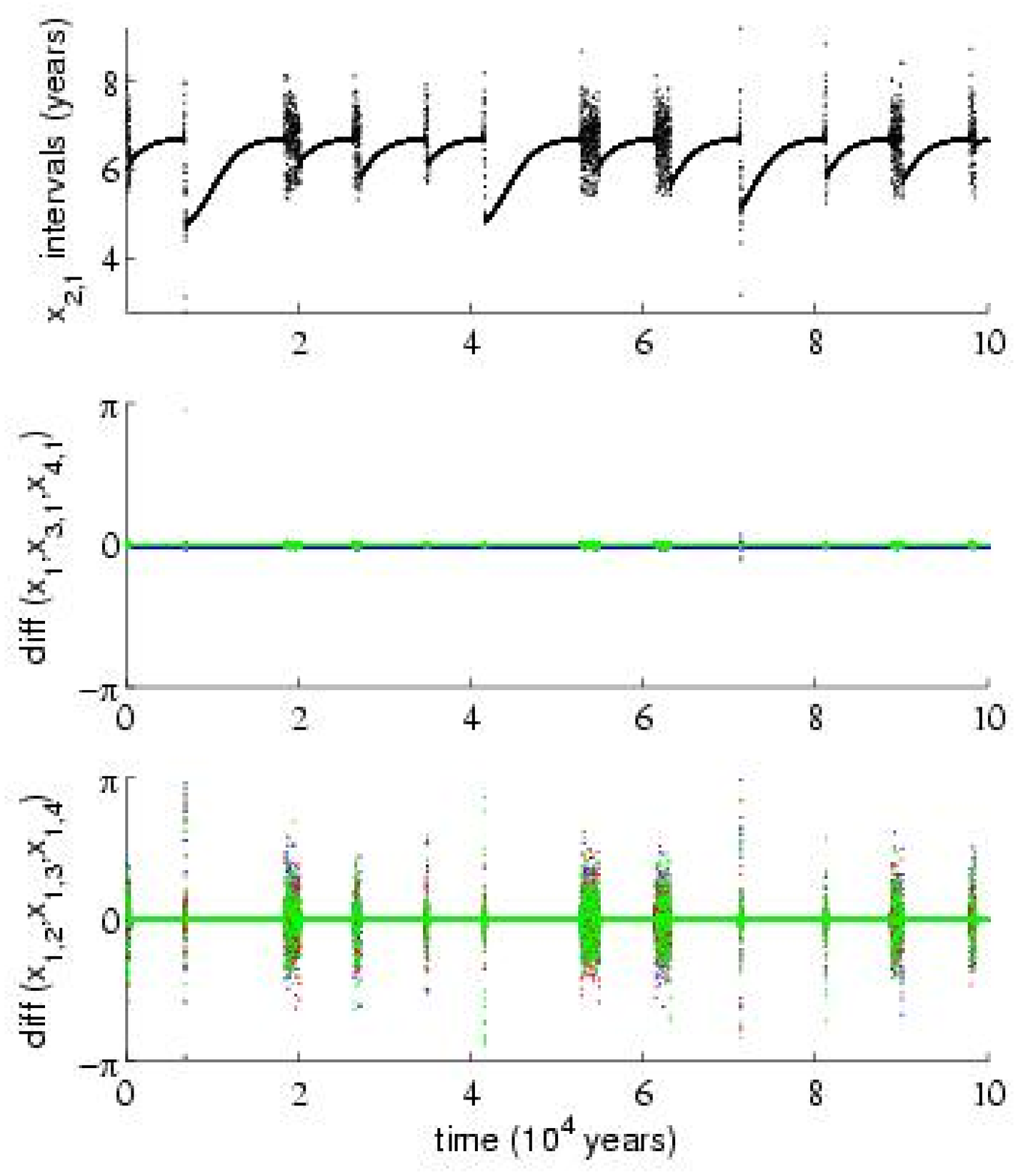} \\
\caption{\label{fig:autonomous phase differences} Phase difference analysis of time series in a chaotic region with $\phi=1.73$ { for autonomous system. (a) Time intervals between local maxima of secondary infective group $x_{2,1}$. 
(b) Phase differences between compartments currently infected with serotype 1.  Primary infective $x_1$ and secondary infectives $x_{3,1}$ and $x_{4,1}$ are compared to $x_{2,1}$.  Maxima occur in phase. (c) Comparison between compartments currently infected with different serotypes.  Secondary infectives  $x_{1,2}$,  $x_{1,3}$, and  $x_{1,4}$ are compared to  $x_{2,1}$.  Maxima occur out of phase during chaotic intervals.}} 
\end{figure}

In Fig.~\ref{fig:autonomous phase differences}, we compare the relative
phases of infective groups for $\phi=1.73$. For the secondary infective group $x_{2,1}$, we plot the
inter-maximum intervals in years in Fig.~\ref{fig:autonomous phase
differences}(a). Notice that during the non-chaotic times, the oscillation
intervals grow slowly, until they begin to vary in an irregular manner
during the chaotic phase. In panels (b) and (c), a direct comparison
between $x_{2,1}$ and the other groups is
plotted using the phase differencing equation $\Psi_{ZY}(\tau_{m})$,
normalized between $-\pi$ and $\pi$. In panel (b), all other
infectives who have serotype 1 as the current infection are
practically in-phase with group $x_{2,1}$. In contrast, in panel (c),
all those having serotype 1 as the primary infection, and currently a
different serotype as the secondary infection, lose synchronization when
the dynamics exhibits chaotic behavior. During the slow buildup phase,
however, the groups are still synchronized. Similar desynchronization 
during chaotic time series occurs for the other
primary and secondary infectives (not pictured here).

\begin{figure}
 \hspace*{-3.2in} (a) \\ \vspace*{-0.15in}

\includegraphics[%
 width=2.25in,
 keepaspectratio]{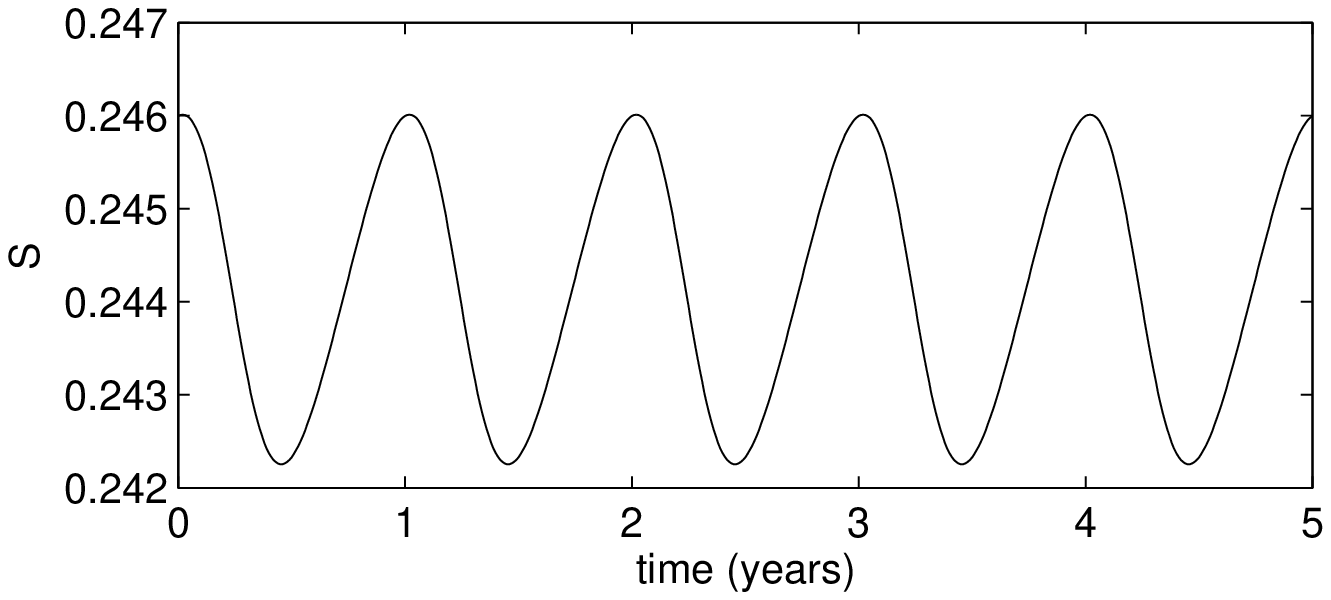} \\
 \hspace*{-1.3in} (b) \hspace*{1.7in} (c) \\ \vspace*{-0.05in}

\includegraphics[%
 width=2.5in,
 keepaspectratio]{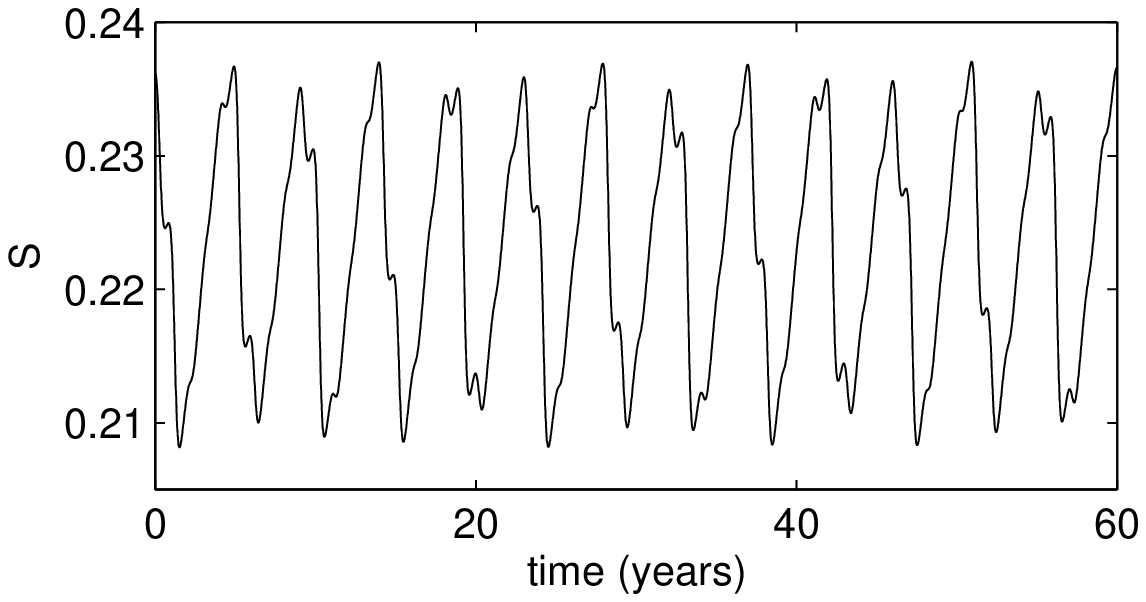} \hspace*{-0.5in}
\includegraphics[%
 width=1.25in,height=1in]{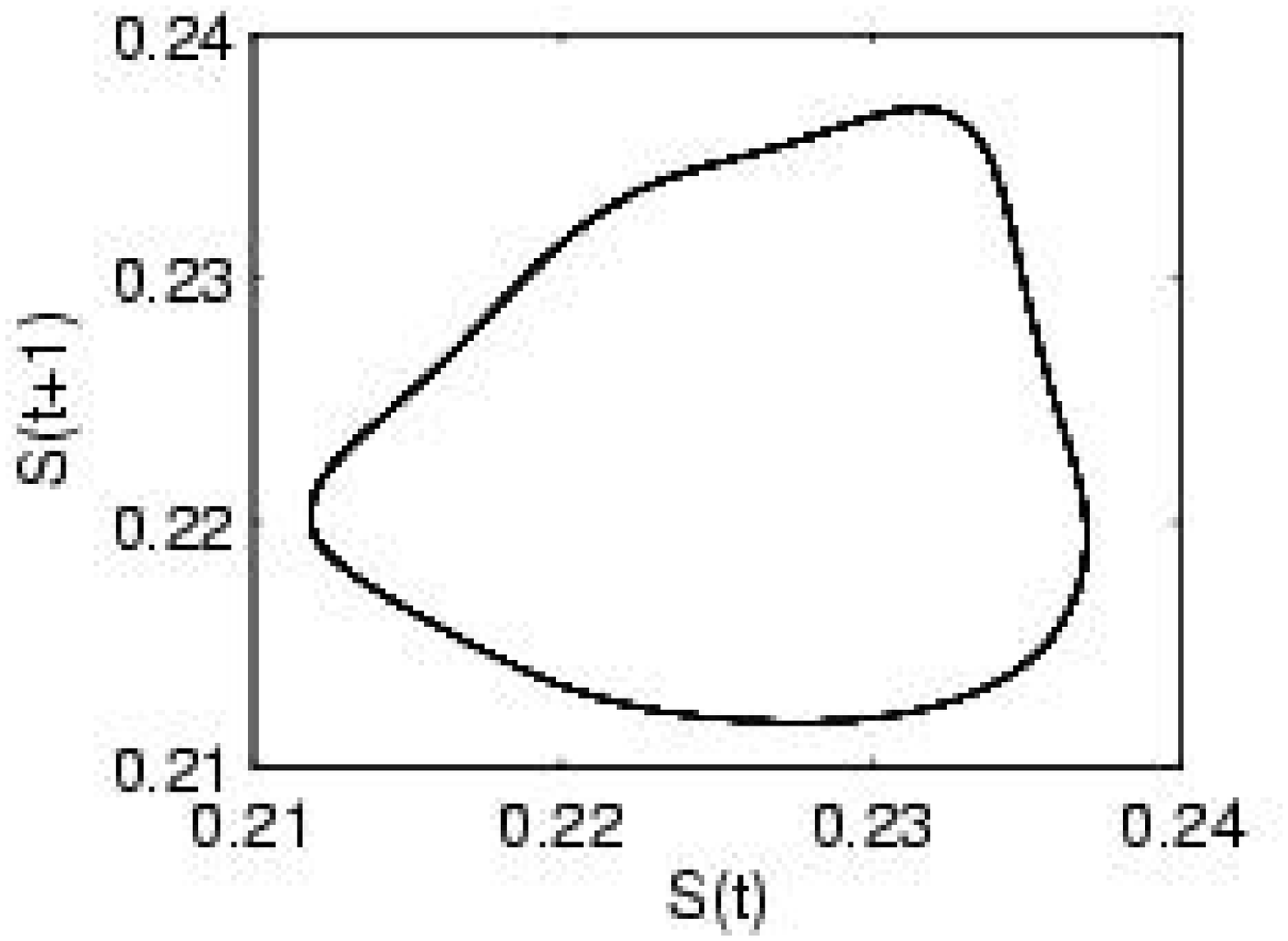} 

\caption{\label{fig:Driven_composite} Dynamics of seasonally driven ADE model, where $\beta_{0}=200$, $\sigma=100$, $\mu=0.02$, and the forcing amplitude $\beta_{1}=0.05$. (a) Periodic time series of susceptible population, for the ADE factor $\phi=$1.5. (b) Quasi-periodic time series of susceptibles, for $\phi=1.7244$. (c) Projected time series sampled at intervals of one year for the susceptible population, showing the quasi-periodicity (same parameters as in (b)). }
\end{figure}

\subsection{Seasonally driven case}

Although the analysis suggests that chaos is responsible for the observed lag between serotypes, one could argue that since there is a seasonal component to the disease, adding a periodic forcing term should synchronize the serotypes, even when they are chaotic. To address this issue, we modified the model to include a contact rate that modulates with a period of one year; i.e., $\beta(t)=\beta_{0}(1+\beta_{1}\cos(2\pi t))$, where $\beta_{1}$ is the forcing amplitude.  ($\beta_1 = 0.05$ was used in this study, but similar behavior is observed for other forcing amplitudes.) The contact rate prefactor, $\beta_{0}=200$, is constant as before. Analogous to the Hopf bifurcation in the autonomous system, bifurcation onto a torus occurs at $\phi_c = 1.7243$. For an ADE factor below $\phi_c$, we observe periodic behavior, as shown in Fig.~\ref{fig:Driven_composite}(a), while for an ADE factor just above $\phi_c$, we find quasi-periodic behavior, plotted in Figs.~\ref{fig:Driven_composite}(b) and (c). In panel (c), the time series of susceptibles was sampled at the forcing period and plotted as successive iterates to show a cross-section of the torus. In both periodic and quasi-periodic cases, the serotypes are all in phase, and there is no desynchronization. However, for higher ADE, we find that the driven system becomes chaotic, and there is desynchronization.

\begin{figure}
\hspace*{-3in} (a) \\
\vspace*{0.9in} \hspace*{-3in} (b) \\
\vspace*{-1.25in} 

\includegraphics[%
 width=2.75in]{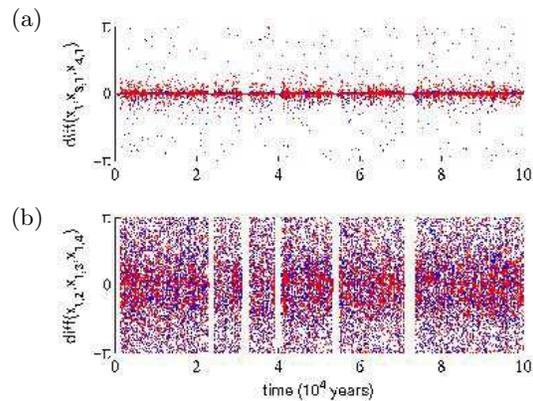} 
\caption{\label{cap:periodically_driven desynch} 
{Chaotic phase desynchronization in periodically driven system. The ADE factor is $\phi=1.74$. 
(a) Phase differences between compartments currently infected with serotype 1.  Primary infective $x_1$ and secondary infectives $x_{3,1}$ and $x_{4,1}$ are compared to $x_{2,1}$.  Maxima usually occur in phase. (b) Comparison between compartments currently infected with different serotypes.  Secondary infectives  $x_{1,2}$,  $x_{1,3}$, and  $x_{1,4}$ are compared to  $x_{2,1}$.  Maxima occur out of phase. (Windows of synchrony occur during the oscillatory regions that have a slowly growing envelope.)}}
\end{figure}

\begin{figure}[t]
\includegraphics[%
 width=2.25in,
 keepaspectratio]{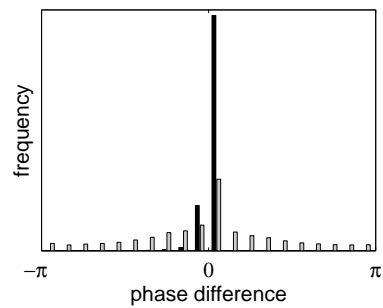} 
\caption{\label{cap:Combined Phase Histo} A histogram plot showing the statistics of the phase differences between secondary infections and primary infections from Fig.~\ref{cap:periodically_driven desynch}. { Black bars:  frequency of phase differences for compartments currently infected with serotype 1 (data from Fig.~\ref{cap:periodically_driven desynch}(a)), gray bars:  frequency of phase differences for compartments currently infected with different serotypes (data from Fig.~\ref{cap:periodically_driven desynch}(b)).}}
\end{figure}

Figure \ref{cap:periodically_driven desynch} shows the phase differences 
between the $x_{2,1}$ secondary infective and other infectives, for an 
ADE factor of $\phi=1.74$ and forcing amplitude $\beta_{1}=0.05$, where 
the solution is chaotic. Notice that in the top panel, where the phase differences are for other compartments currently infected with serotype 1, there is phase synchrony on average when 
compared to the case where the secondary infections are from a different 
serotype (second panel). Although the phase synchrony is not as good as in 
the autonomous case in Fig.~\ref{fig:autonomous phase differences}, we can 
get a statistical measure showing how on average the phase locking 
compares by computing a histogram of both cases. This is shown in 
Fig.~\ref{cap:Combined Phase Histo}, where the grey bars correspond to 
phase differences of Fig.~\ref{cap:periodically_driven desynch}(b), and 
the black bars correspond to the data from \ref{cap:periodically_driven 
desynch}(a). Notice that when comparing primary infections of serotype 1 
to secondary infections that currently have serotype 1, there is a strong 
phase locking component on average. 

\section{Conclusions}

{ We have derived and  analyzed the dynamics of  a model for multi-strain diseases with
  antibody-dependent enhancement.  The model for secondary infections,
  which includes ADE as a parameter, adds  a new wrinkle to models
  of the SIR type. In previous studies of single strain models that do
  not include  environmental forcing, the endemic equilibrium is the
  only possible  stable state. That is, there are no bifurcations which give
  rise to dynamics exhibiting regular or irregular outbreaks. In
  contrast, by modeling  the effect of ADE as an increase in
  infectivity of secondary infections, we see both analytically and
  numerically that periodic outbreaks appear at a critical ADE
  value. Moreover, the analysis 
  reveals exactly how the period of oscillations depends on the ADE
  parameter near the bifurcation point. The range of periods predicted
  for the parameters used in our computations appear to agree  well with those
  observed in  the data in Fig.~\ref{fig:Nisalak data} in the
  appendix. 

When the ADE factor increases above
  a threshold, the system's behavior is chaotic, and outbreaks of
  different strains occur asynchronously.  This observation
  corresponds qualitatively with epidemiological data on asynchronous
  outbreaks of dengue fever (see Appendix).}  Seasonal forcing, thought to be a primary driver for the observed
oscillations in the different strains, is typically believed to
disrupt any out-of-phase behavior in the dynamics and force the entire
system to lock on the period of the forcing. However, in our
preliminary study, we find that this is not the case. Phase desynchronization between serotypes occurs even in the seasonally forced case.

However, there
exists a specific relationship between the primary and secondary
infections. Specifically,  we have observed that although the different
serotypes desynchronize when the solutions are 
chaotic, there is surprising structure in the peak outbreaks of the
serotypes when 
comparing the appropriate secondary infectives to the appropriate primary
infectives.  { Although} there is no vaccine currently available for all
serotypes, the 
results here point to potential new methods of analysis and monitoring of
multi-strain diseases.  In the field, the majority of the cases reported are secondary infections. 
Therefore, by observing a small percentage of the incidence in the secondary 
infections of one serotype, synchronization would imply that the data is 
representative of the general behavior of all the groups infected with that 
serotype, including those with only a primary infection. {Further global analysis techniques based on center
manifold methods can be used to explain the synchronization of particular primary and secondary infectives
when the time series becomes chaotic; this approach is the subject of further study}.  

Research was supported by the Office of Naval Research and the Center for
Army Analysis. LB and MM were supported by the National Science Foundation under Grants DMS-0414087 and
CTS-0319555. LBS is currently a National Research Council post doctoral fellow.
DATC and DSB were supported by the National Institutes of Health under
Grant U01GM070708 and by the National Oceanic and Atmospheric
Administration under Grant NA04OAR4310138.

\appendix

\section{Epidemiological data}

\begin{figure*}[t]
\includegraphics[%
 width=7in,
 keepaspectratio]{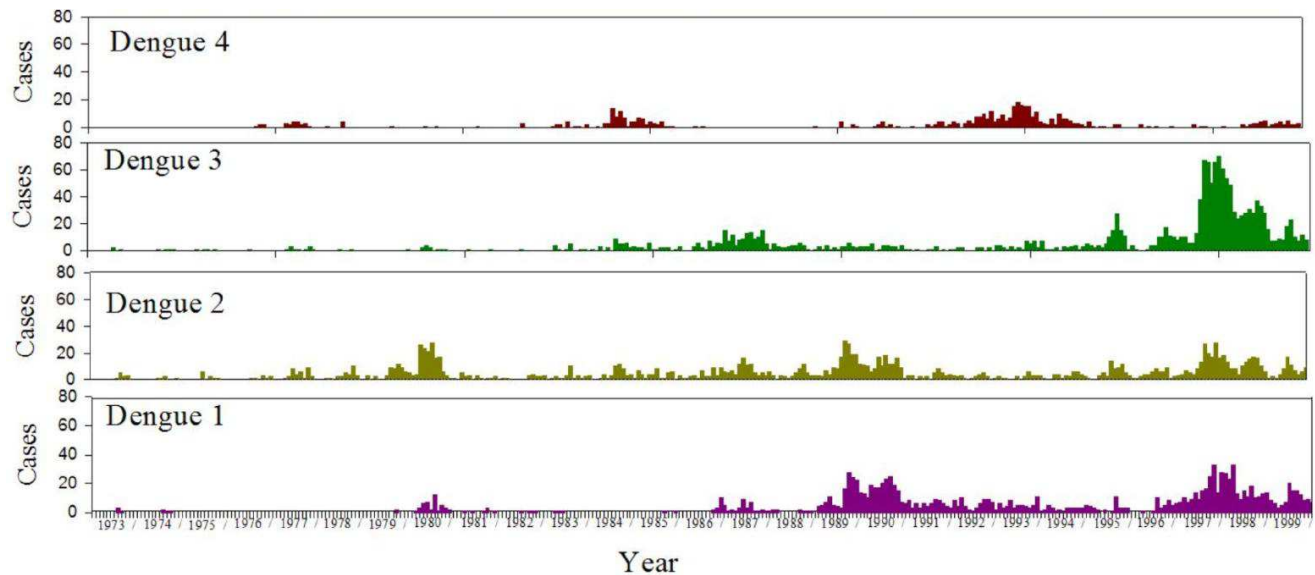} 
\caption{\label{fig:Nisalak data} {Frequency of detection  of each of
    the four Dengue virus types per month at the Queen Sirikit
    National Institute for Child Health from 1973-1999.  Reprinted
    from \cite{NisalakENKTSBHIV03}.}} 
\end{figure*}  


Fig.~\ref{fig:Nisalak data}, reprinted from \cite{NisalakENKTSBHIV03},
shows the frequency of detection of each of the four dengue types at one
hospital in Bangkok, Thailand, over a continuous 27 year period of
monitoring. Infecting serotypes were defined by isolation of replication-
competent virus and/or detection of viral genome in peripheral blood. (It
should be noted that serological measurements were performed for only a
fraction of all dengue cases.)  Observe that peaks of the dengue virus
types are asynchronous.



\end{document}